\documentclass[conference]{IEEEtran}
\IEEEoverridecommandlockouts

\usepackage{cite}
\usepackage{amsmath,amssymb,amsfonts}
\usepackage{graphicx}
\usepackage{textcomp}
\usepackage{xcolor}
\usepackage{array}
\usepackage{url}
\usepackage{multirow}
\usepackage{booktabs}
\usepackage{makecell}
\usepackage{float}
\usepackage{bbm}
\usepackage{tikz}
\usetikzlibrary{arrows.meta,positioning}
\usepackage[caption=false,font=footnotesize]{subfig}
\usepackage[normalem]{ulem}
\usepackage{booktabs}
\usepackage{algorithm}
\usepackage{algpseudocode}

\def\BibTeX{{\rm B\kern-.05em{\sc i\kern-.025em b}\kern-.08em
    T\kern-.1667em\lower.7ex\hbox{E}\kern-.125emX}}

\begin{document}

\title{Channel-Chart–Driven Antenna Activation for Energy-Efficient Cell-Free Massive MIMO}

\author{
    \IEEEauthorblockN{
        Yara Kafa, Luc Le Magoarou, Joumana Farah, Matthieu Crussière
    }
    \IEEEauthorblockA{
        Univ Rennes, INSA Rennes, CNRS, IETR-UMR 6164, Rennes, France
    }
}

\maketitle

\begin{abstract}
Cell-free massive MIMO (CF-mMIMO) can provide uniformly high system performance across the coverage area through distributed cooperation. However, keeping all antennas active leads to high circuit and fronthaul power consumption. This paper studies energy-efficient uplink reception through \emph{antenna activation}, where only a subset of distributed antennas is selected to serve the users. To enable network-wide activation with low signaling overhead, each access point (AP) maps its locally estimated channel state information (CSI) into a low-dimensional \emph{channel-chart (CC) embedding} and forwards this representation to the central processing unit (CPU) instead of the full CSI. Based on the received embeddings, we formulate a binary antenna-activation problem for network EE maximization and propose three low-overhead strategies: (i) direct embedding-domain activation using a proxy EE metric without CSI reconstruction, (ii) nearest chart-point CSI approximation, which estimates CSI from a reference chart database for greedy EE optimization, and (iii) decoder-based CSI reconstruction using a pre-trained autoencoder.
The proposed approaches further enable a systematic comparison of antenna activation techniques in terms of achieved performance, computational complexity, and signaling overhead. Numerical results show substantial EE gains over the all-active baseline, with only a minor degradation in sum spectral efficiency and a significant reduction in signaling overhead. Moreover, the proposed greedy framework achieves EE performance close to that of the simulated annealing baseline, confirming its effectiveness for low-complexity antenna activation.
\end{abstract}


\section{Introduction}
CF-mMIMO  (Cell-Free Massive MIMO) consists of a large set of geographically distributed antennas that serve users collaboratively, with their received uplink signals being coherently combined and processed at a central processing unit (CPU)~\cite{7827017,demir2021foundations}. While this architecture provides strong macro-diversity and robustness, it requires operating a large number of radio-frequency (RF) chains and fronthaul links, leading to substantial circuit and fronthaul power consumption~\cite{8097026}. Consequently, keeping all distributed antennas active may yield only limited additional reception benefits while further increasing the network power cost. This motivates adaptive \emph{network-wide antenna activation}, where only the most useful antennas are dynamically selected for uplink reception to improve energy efficiency while retaining the macro-diversity gains of CF-mMIMO.
Energy-efficient antenna selection and activation have been widely studied in conventional MIMO, massive MIMO, and XL-MIMO systems~\cite{7100915,9241062,9187980}, using channel-dependent selection rules based on instantaneous CSI, imperfect CSI, long-term channel statistics, or learning-based optimization, and mainly focusing on centralized or co-located array deployments. In CF-mMIMO, related designs have mainly considered power control, AP selection, pilot assignment, and/or structured cooperation, generally relying on centralized optimization or coordination based on CSI-derived channel statistics~\cite{8097026,9174860}. While effective, such approaches typically operate at the AP-UE association or cooperation level and do not address fine-grained, network-wide per-antenna activation from compact CSI representations, which is important for scalable dense deployments. Although joint antenna activation and power allocation under wireless fronthaul constraints has recently been investigated~\cite{10942847}, that approach remains centralized and CSI/model-driven, requiring AP-UE access-channel statistics and wireless-fronthaul CSI to enforce SINR and fronthaul-capacity constraints. It therefore does not address antenna activation from compact learned CSI representations. This motivates \emph{network-wide, explicit antenna activation} based on channel-charting (CC) embeddings, enabling centralized activation decisions with reduced signaling overhead.
To address this gap, we exploit low-dimensional CC embeddings of locally estimated CSI~\cite{9526758}. Each AP, equipped with multiple antennas, compresses its instantaneous channel observations into a compact CC embedding, and forwards them to the CPU, thereby drastically reducing the fronthaul signaling overhead. The \textbf{contributions of this work} are as follows:
we develop a greedy EE-maximization framework for scalable network-wide antenna activation under low signaling overhead, while closely approaching the performance of a full-CSI-based antenna-activation benchmark. Within this framework, we further propose three low-overhead activation strategies: a direct embedding-domain method that performs per-antenna activation across all APs directly in the compressed CC space by maximizing a proposed proxy EE metric, avoiding CSI reconstruction; a nearest chart-point CSI approximation strategy that estimates approximate CSI by matching the received CC embedding to a calibrated reference chart point and then performs greedy maximization of the exact EE objective; and a decoder-based strategy that instead reconstructs channel estimates from CC embeddings using a pre-trained autoencoder for greedy EE optimization. All methods are evaluated using realistic synthetic channels generated with the Sionna ray-tracing framework~\cite{sionna}, where directive three-dimensional antenna radiation patterns with measured non-isotropic angular responses~\cite{batel2019superdirective,tornese2024endfire} are integrated into the ray-tracing setup. The antennas are arranged in a cylindrical array, producing directive beams that jointly cover the $360^\circ$ azimuth range, thereby creating spatially selective channel conditions that support meaningful antenna-activation decisions across the distributed CF-mMIMO network. These methods enable a systematic comparison between direct compressed-domain activation and CSI-based activation with different reconstruction fidelities, thereby allowing us to evaluate the tradeoffs among activation performance, computational complexity, and required fronthaul signaling overhead. To the best of our knowledge, this is the first work to investigate network-wide, per-antenna activation in CF-mMIMO driven by CC representations.

\section{System Model}
\subsection{System Description}
We consider the uplink of a user-centric CF-mMIMO system with $L$ distributed APs serving $K$ single-antenna users (UEs). Each AP $\ell\in\{1,\ldots,L\}$ is equipped with $N$ antennas, yielding a total of $M=LN$ antennas, and is connected to a CPU via fronthaul links. At each AP, the antennas are arranged in a uniform cylindrical array composed of $N_{\mathrm{st}}$ vertically stacked stages, each containing the same number of antennas uniformly distributed around the cylindrical aperture, so that the full array jointly covers the $360^\circ$ azimuth range. The system operates in Time Division Duplexing (TDD) mode over a subband of bandwidth $B_{\mathrm{sub}}$ consisting of $S$ subcarriers, where CSI is locally estimated at the APs using $\tau_p$ uplink pilot symbols. The effective uplink channel between user~$k$ and AP~$\ell$, incorporating the spatial selectivity induced by the directive three-dimensional antenna radiation patterns, is denoted by $\mathbf{h}_{k\ell}\in\mathbb{C}^{N}$, and the stacked global channel of user~$k$ is 
$\mathbf{h}_k \triangleq [\mathbf{h}_{k1}^\top \ \cdots \ \mathbf{h}_{kL}^\top]^\top \in \mathbb{C}^{M}$. To limit signaling and processing complexity, user-centric dynamic cooperation clustering (DCC) at the AP level is adopted \cite{9174860}. Let $\mathcal{M}_k$ and $\mathcal{A}_\ell$ denote the sets of APs serving user~$k$ and UEs served by AP~$\ell$, respectively, obtained using the AP-UE association algorithm from~\cite{9174860}, with $|\mathcal{A}_\ell|\leq\tau_p$. The resulting association matrix $\mathbf{D}\in\{0,1\}^{L\times K}$ has entries $D_{\ell k}=1$ if AP~$\ell$ serves user~$k$, and $0$ otherwise, and the corresponding DCC matrix for user~$k$ is $\mathbf{D}_k \triangleq \mathrm{diag}(D_{1k}\mathbf{I}_N,\ldots,D_{Lk}\mathbf{I}_N)\in\mathbb{C}^{M\times M}$. 

\subsection{Uplink Signal Model and Antenna Activation}
The received uplink signal at AP~$\ell$ is
$\mathbf{r}_\ell=\sum_{k=1}^{K}\sqrt{p_k}\mathbf{h}_{k\ell}x_k+\mathbf{n}_\ell\in\mathbb{C}^{N}$,
where $x_k$ denotes the unit-power data signal of user~$k$, $p_k$ is the uplink transmit power, 
identical for all users, and
$\mathbf{n}_\ell\sim\mathcal{CN}(\mathbf{0},\sigma^2\mathbf{I}_N)$ the additive noise.
Antenna activation at AP~$\ell$ is modeled by a binary vector
$\boldsymbol{\alpha}_\ell\in\{0,1\}^{N}$, whose $n$th entry $\alpha_{\ell,n}=1$ if antenna~$n$ of AP~$\ell$ is active and 0 otherwise. Collecting all per-AP activation vectors yields the global vector
$\boldsymbol{\alpha}\triangleq[\boldsymbol{\alpha}_1^\top\ \cdots\ \boldsymbol{\alpha}_L^\top]^\top\in\{0,1\}^{M}$,
and the corresponding global activation matrix
$\mathbf{\Lambda}\triangleq \mathrm{diag}(\boldsymbol{\alpha})\in\{0,1\}^{M\times M}$, with per-AP activation matrix
$\mathbf{\Lambda}_\ell\triangleq\mathrm{diag}(\boldsymbol{\alpha}_\ell)$.

\subsection{Uplink Combining over Activated Antenna Ports}

Once optimized at the CPU, the activation vector $\boldsymbol{\alpha}$ is fed back to the APs, where each AP applies its local matrix $\mathbf{\Lambda}_\ell$ so that uplink detection, baseband processing, and fronthaul forwarding involve only the selected antenna ports. MMSE-based combining is then adopted for uplink detection, considering both centralized and distributed implementations.

\textbf{Centralized partial MMSE (C\_MMSE):}
In the centralized scheme, the CPU performs uplink combining and thus requires the instantaneous CSI associated with the active antennas. To limit the combining complexity, partial MMSE (P-MMSE) reception is adopted by accounting only for the interference from users sharing at least one serving AP, as in~\cite{demir2021foundations}.
Define
\(
\mathcal{S}_k \triangleq \{ i : \mathbf{D}_k \mathbf{D}_i \neq \mathbf{0} \}.
\)
The combining vector for user~$k$ is \begin{equation} \mathbf{w}_k = \left( \sum_{i \in \mathcal{S}_k} p_i \mathbf{D}_k \mathbf{\Lambda} \mathbf{h}_i \mathbf{h}_i^H \mathbf{\Lambda} \mathbf{D}_k + \sigma^2 \mathbf{D}_k \mathbf{\Lambda} \mathbf{D}_k \right)^{-1} \mathbf{D}_k \mathbf{\Lambda} \mathbf{h}_k . \end{equation}

\textbf{Distributed MMSE (D\_MMSE):}
In the distributed scheme, each AP applies local MMSE detection, while the CPU performs equal-weight combining to the locally detected signals, as in~\cite{demir2021foundations}.
The local combining vector for user~$k$ at AP~$\ell$ is
\begin{equation} \mathbf{w}_{k\ell} = \left( \sum_{i \in \mathcal{A}_\ell} p_i \mathbf{\Lambda}_\ell \mathbf{h}_{i\ell} \mathbf{h}_{i\ell}^H \mathbf{\Lambda}_\ell + \sigma^2 \mathbf{\Lambda}_\ell \right)^{-1} \mathbf{\Lambda}_\ell \mathbf{h}_{k\ell} D_{\ell k}. \end{equation}

\textbf{Unified Uplink SINR:}
Define the global combining vector $\mathbf{w}_k \in \mathbb{C}^{M}$ as the P-MMSE combiner in the centralized case, or as
\(
\mathbf{w}_k \triangleq [\mathbf{w}_{k1}^\top \ \cdots \ \mathbf{w}_{kL}^\top]^\top
\)
in the distributed case, with $\mathbf{w}_{k\ell}=\mathbf{0}$ for $\ell \notin \mathcal{M}_k$. The SINR of user~$k$ is
\begin{equation}
\gamma_k(\boldsymbol{\alpha}) =
\frac{
p_k \left| \mathbf{w}_k^{H} \mathbf{D}_k \mathbf{\Lambda} \mathbf{h}_k \right|^2
}{
\sum_{i \neq k}
p_i \left| \mathbf{w}_k^{H} \mathbf{D}_k \mathbf{\Lambda} \mathbf{h}_i \right|^2
+ \sigma^2 \mathbf{w}_k^{H} \mathbf{D}_k \mathbf{\Lambda} \mathbf{D}_k \mathbf{w}_k
}.
\end{equation}

\section{Energy Efficiency Problem Formulation}
We consider uplink EE maximization under antenna activation (AAc), formulated as
\begin{equation}
\underset{\boldsymbol{\alpha}\in\{0,1\}^{M}}{\text{maximize}}\quad
\mathrm{EE}(\boldsymbol{\alpha})
\triangleq
\frac{R(\boldsymbol{\alpha})}{P_{\mathrm{tot}}(\boldsymbol{\alpha})},
\label{eq:EE_problem}
\end{equation}
where the achievable uplink rate of user~$k$ is defined as
$R_k(\boldsymbol{\alpha})
=
B_{\mathrm{sub}}\log_2\!\bigl(1+\gamma_k(\boldsymbol{\alpha})\bigr)$.
The corresponding uplink sum rate is
$R(\boldsymbol{\alpha}) =
\sum_{k=1}^{K} R_k(\boldsymbol{\alpha})$. AAc is constrained by the DCC association, such that antennas can be activated only at APs serving at least one user. Let $\delta_\ell \triangleq \mathbbm{1}\!\left(\sum_{k=1}^{K} D_{\ell k} > 0\right)$ denote an indicator of whether AP~$\ell$ serves at least one user. The total uplink power consumption is modeled following~\cite{9849036}, adapted to the antenna activation context, as
\begin{equation}
P_{\mathrm{tot}}(\boldsymbol{\alpha})
=
\sum_{k=1}^{K}\frac{p_k}{\nu_k}
+
\sum_{\ell=1}^{L}
\delta_\ell\, P_{\mathrm{tc}}
\sum_{n=1}^{N}\alpha_{\ell,n}
+
\sum_{\ell=1}^{L} P_{\mathrm{fh},\ell}(\boldsymbol{\alpha}),
\end{equation}
where $\nu_k\in(0,1]$ is the power-amplifier efficiency at user~$k$ and $P_{\mathrm{tc}}$ is the internal circuit and processing power per active antenna. The fronthaul power consumption at AP~$\ell$ is $P_{\mathrm{fh},\ell}(\boldsymbol{\alpha})
=
P_{\mathrm{fix},\ell}
+
P_{\mathrm{ft},\ell}
\sum_{k=1}^{K} D_{\ell k}\, R_k(\boldsymbol{\alpha})$,
where $P_{\mathrm{fix},\ell}$ denotes the fixed fronthaul power and
$P_{\mathrm{ft},\ell}$~(W/(bit/s)) denotes the traffic-dependent power coefficient. Due to the binary activation variables and the coupling among antenna selection, multiuser interference, and MMSE combining, problem~\eqref{eq:EE_problem} is a nonconvex combinatorial fractional program. In the sequel, low-overhead CC-based strategies are developed to efficiently approximate solutions to~\eqref{eq:EE_problem}.

\section{Antenna-Activation Strategies Under Low-Overhead CSI Exchange}
This section presents three CC-based antenna-activation strategies for solving the EE-maximizing AAc problem in~\eqref{eq:EE_problem}, whose objective is to determine the network-wide binary activation vector $\boldsymbol{\alpha}$. Instead of relying on full high-dimensional CSI at the CPU, these strategies use the channel-chart embeddings forwarded by the APs as input to the AAc procedure. This common CC-based formulation enables a systematic comparison among the three strategies in terms of the computational complexity required to compute $\boldsymbol{\alpha}$ and the signaling overhead toward the CPU.

\subsection{Channel-Chart Generation and Embedding Exchange}
CC provides low-dimensional representations of locally estimated CSI that preserve neighborhood relationships (see~\cite{8444621,9526758,10051826} for the complete formulation). In this work, CC is used to enable AAc with reduced fronthaul signaling.

\subsubsection{Offline CC Construction}
During an offline calibration phase, each AP independently constructs a CC from locally estimated uplink CSI collected from a set of calibration users. Let $\mathbf{h}_{\ell}^{(i)}\in\mathbb{C}^{N}$ denote the effective channel of calibration user~$i$ at AP~$\ell$, obtained by averaging the per-subcarrier channel estimates over the $S$ subcarriers in the considered subband. This averaging mitigates frequency-dependent variations and yields a geometrically robust representation $\mathbf{h}_{\ell}^{(i)}$, which is essential for learning a stable and meaningful CC. Following~\cite{9526758}, the phase-insensitive similarity between calibration channels $i$ and $j\neq i$ is defined as $s_{\ell}(i,j)\triangleq \frac{\left|\mathbf{h}_{\ell}^{(i)H}\mathbf{h}_{\ell}^{(j)}\right|}{\|\mathbf{h}_{\ell}^{(i)}\|\,\|\mathbf{h}_{\ell}^{(j)}\|}\in[0,1]$, based on which the corresponding channel distance is given by $d_{\ell}(i,j)=\sqrt{2-2\,s_{\ell}(i,j)}$. Isomap, a nonlinear dimensionality reduction method~\cite{tenenbaum2000isomap}, is then applied to the resulting pairwise distance matrix, whose entries contain the channel distances $d_{\ell}(i,j)$ between all calibration-user pairs $i\neq j$, to obtain the reference chart coordinates $\{\mathbf{z}_{\ell}^{(i)}\in\mathbb{R}^{d}\}$, where $d$ denotes the CC dimension. This mapping places calibration users with similar channel characteristics at nearby chart points. The CC construction is performed offline and reused during online AAc operation.

\subsubsection{Forward Charting and Embedding Exchange}
During online operation, each AP maps newly observed CSI onto the precomputed CC through forward channel charting.
Specifically, for user~$k$ served by AP~$\ell$, the instantaneous channel vector $\mathbf{h}_{k\ell}$ is compared with the stored calibration-user channels using the phase-insensitive similarity measure $s_{\ell}(k,i)\triangleq \frac{\left|\mathbf{h}_{k\ell}^{H}\mathbf{h}_{\ell}^{(i)}\right|}{\|\mathbf{h}_{k\ell}\|\,\|\mathbf{h}_{\ell}^{(i)}\|}$, for $i=1,\ldots,N_{\mathrm{c}}$, where $N_{\mathrm{c}}$ denotes the number of calibration users at AP~$\ell$.
The $K_{\mathrm{nn}}$ most similar calibration-user samples are selected, and their corresponding chart coordinates are combined through weighted interpolation to obtain the low-dimensional embedding $\mathbf{z}_{k\ell}\in\mathbb{R}^{d}$.
This feed-forward procedure~\cite{9526758,10051826} has linear complexity in the number of calibration users and relies on only local CSI.
The resulting embeddings $\mathbf{z}_{k\ell}$ are then sent to the CPU, enabling centralized AAc with substantially reduced fronthaul signaling compared with full CSI exchange.

\subsection{Antenna-Activation Strategies Based on Channel Charts}
Given the CC embeddings available at the CPU, we propose three low-overhead AAc strategies for solving problem ~\eqref{eq:EE_problem}, each exploiting the embeddings in a different way.
\subsubsection{Direct Embedding-Domain Antenna Activation (DD)}
This strategy performs AAc in the CC domain without reconstructing instantaneous CSI at the CPU. The embeddings are instead used to define a proxy EE metric that guides antenna selection with minimal computational and signaling overhead.

\paragraph{Embedding-to-Channel-Power Prediction}
The approach relies on the observation that users with correlated channel vectors at an AP exhibit similar per-antenna channel power profiles. Since the CC is constructed to preserve this local similarity between neighboring users, proximity in the embedding space enables channel-power prediction without requiring instantaneous CSI at the CPU. During the offline phase, each AP~$\ell$ constructs a CPU-stored calibration lookup table,
$\{(\mathbf{z}^{(i)}_\ell,\mathbf{u}^{(i)}_\ell)\}_{i=1}^{N_\ell^{\mathrm{c}}}$,
where $i$ indexes calibration users at AP~$\ell$,
$\mathbf{z}^{(i)}_\ell\in\mathbb{R}^{d}$ is a calibration-user embedding, and
$\mathbf{u}^{(i)}_\ell=[|h^{(i)}_{\ell,1}|^2,\ldots,|h^{(i)}_{\ell,N}|^2]^\top$
is the corresponding per-antenna channel power vector.
During online operation, for a user~$k$ with embedding $\mathbf{z}_{k\ell}$, the CPU searches this lookup table to identify the set $\mathcal{N}_{k\ell}$ of nearest calibration users. The predicted per-antenna channel power vector is obtained as
$\widehat{\mathbf{u}}_{k\ell}
=
\sum_{i\in\mathcal{N}_{k\ell}}
\omega^{(i)}_{k\ell}\,\mathbf{u}^{(i)}_\ell$,
where the weights are given by
\begin{equation}
\omega^{(i)}_{k\ell}
=
\frac{\|\mathbf{z}_{k\ell}-\mathbf{z}^{(i)}_\ell\|_2^{-1}}
{\sum_{j\in\mathcal{N}_{k\ell}}\|\mathbf{z}_{k\ell}-\mathbf{z}^{(j)}_\ell\|_2^{-1}} .
\label{eq:utility_weights}
\end{equation}
Stacking across APs yields the predicted channel-power matrix $\widehat{\mathbf{U}}\in\mathbb{R}_+^{K\times M}$.

\paragraph{Embedding-Based User-Similarity Metric}
During inference, the average similarity of user~$k$ at AP~$\ell$ is computed as
$\rho_{k\ell}=\frac{1}{K-1}\sum_{i\neq k}\left(1-\frac{1}{2}\|\mathbf{z}_{k\ell}-\mathbf{z}_{i\ell}\|_{2}^{2}\right)$, where $\|\mathbf{z}_{k\ell}-\mathbf{z}_{i\ell}\|_{2}$ denotes the normalized Euclidean distance between the chart embeddings of users $k$ and $i$.
This metric reflects how closely user~$k$ is clustered with the other users in the chart space at AP~$\ell$ and is later used through the factor $(1-\rho_{k\ell})$ to penalize such contributions in the proxy utility.

\paragraph{Embedding-Based Proxy SINR and EE}
Let $M_{\mathrm{act}}(\boldsymbol{\alpha})=\sum_{\ell=1}^{L}\sum_{n=1}^{N}\alpha_{\ell,n}$ denote the total number of activated antennas.
Using the predicted channel-power matrix $\widehat{\mathbf{U}}$ and the embedding-based similarity metric $\rho_{k\ell}$, we define the proxy utility of user~$k$ as
\begin{equation} \label{eq:utility}
\widehat{\eta}_k(\boldsymbol{\alpha})
=
\sum_{\ell=1}^{L}
\sum_{n=1}^{N}
\alpha_{\ell,n}\widehat{u}_{k\ell,n}
\frac{1-\rho_{k\ell}}{1+\sum_{n'=1}^{N}\alpha_{\ell,n'}} ,
\end{equation}
where $\sum_{n'=1}^{N}\alpha_{\ell,n'}$ denotes the number of antennas activated at AP~$\ell$ under the candidate activation vector $\boldsymbol{\alpha}$. Each term in \eqref{eq:utility} quantifies the desirability of activating antenna~$n$ at AP~$\ell$ for user~$k$.
It rewards antennas with high predicted channel power, 
penalizes AP-user contributions associated with highly clustered users through the factor $(1-\rho_{k\ell})$, and promotes sparse activation at each AP through the normalization by the number of active antennas at that AP.
Based on $\widehat{\eta}_k(\boldsymbol{\alpha})$, the proxy SINR is defined as
\begin{equation}\label{eq:gamma}
\widehat{\gamma}_k(\boldsymbol{\alpha})
=
\frac{p_k}{\sigma^2}\widehat{\eta}_k(\boldsymbol{\alpha})
\frac{M_{\mathrm{act}}(\boldsymbol{\alpha})-K}{M_{\mathrm{act}}(\boldsymbol{\alpha})},
\quad M_{\mathrm{act}}(\boldsymbol{\alpha})>K.
\end{equation}
This proxy captures the effects of the transmit signal-to-noise ratio of user~$k$, the defined proxy utility, and the activated-antenna factor. The term $\tfrac{M_{\mathrm{act}}(\boldsymbol{\alpha})-K}{M_{\mathrm{act}}(\boldsymbol{\alpha})}$ favors activation sets with more antennas than users, reflecting the additional spatial degrees of freedom available for multi-antenna reception.

The proxy $\widehat{\mathrm{EE}}(\boldsymbol{\alpha})$ is obtained by replacing the exact SINR $\gamma_k(\boldsymbol{\alpha})$ in~\eqref{eq:EE_problem} with the proxy SINR $\widehat{\gamma}_k(\boldsymbol{\alpha})$.

\paragraph{Greedy Antenna Activation Using Proxy EE}
AAc is performed using Algorithm~\ref{alg:greedy_compact}. 
A low-overhead solution is obtained by replacing
the true $\mathrm{EE}(\boldsymbol{\alpha})$ with
the proxy $\widehat{\mathrm{EE}}(\boldsymbol{\alpha})$, computed from CC-derived quantities via \eqref{eq:utility} and~\eqref{eq:gamma}.
The candidate and active antenna sets are denoted by $\mathcal{C}$ and $\mathcal{S}$, respectively. 
The algorithm uses a warm start by activating one antenna per user based on the predicted channel powers $\widehat{u}_{k,m}$, and then incrementally expands $\mathcal{S}$ to maximize $\widehat{\mathrm{EE}}(\boldsymbol{\alpha})$. 
Leveraging the empirically observed concave dependence of EE on the number of active antennas, the procedure terminates when no further improvement is achieved.

\begin{algorithm}[t]
\caption{Greedy antenna activation}
\label{alg:greedy_compact}
\begin{algorithmic}[1]
\Statex \textbf{Input:} $\{\mathbf{z}_{k\ell}\}_{k,\ell}$, $\{p_k\}$, $\mathbf{D}$
\Statex \textbf{Output:} $\boldsymbol{\alpha}^\star$, $M_{\mathrm{act}}^\star$
\Statex \textbf{Init:} $\mathcal{C}\!=\!\{1,\ldots,LN\}$, $\mathcal{S}\!\gets\!\emptyset$, $\boldsymbol{\alpha}\!\gets\!\mathbf{0}$,
$M_{\mathrm{act}}\gets K+1$, $\mathrm{EE}_{\max}\!\gets\!-\infty$, $\boldsymbol{\alpha}^\star\!\gets\!\mathbf{0}$

\For{$k=1,\ldots,K$} \Comment{Warm start}
    \State $m^\star\gets\arg\max_{m\in\mathcal{C}} u_{k,m}$  \textcolor{gray}{//$u_{k,m}=|h_{k,m}|^2$}
    \If{$\alpha_{m^\star}=0$}
        \State $\alpha_{m^\star}\gets1$, $\mathcal{S}\gets\mathcal{S}\cup\{m^\star\}$
    \EndIf
\EndFor

\While{true}
    \While{$|\mathcal{S}|<M_{\mathrm{act}}$}
        \State $m^\star\gets\arg\max_{m\in\mathcal{C}\setminus\mathcal{S}} \mathrm{EE}(\boldsymbol{\alpha}\,|\,\alpha_m\!=\!1)$
        \State $\alpha_{m^\star}\gets1$, $\mathcal{S}\gets\mathcal{S}\cup\{m^\star\}$
    \EndWhile
    \State $\mathrm{EE}_{\mathrm{cur}}\gets\mathrm{EE}(\boldsymbol{\alpha})$
    \If{$\mathrm{EE}_{\mathrm{cur}}>\mathrm{EE}_{\max}$}
        \State $\mathrm{EE}_{\max}\gets\mathrm{EE}_{\mathrm{cur}}$, $\boldsymbol{\alpha}^\star\gets\boldsymbol{\alpha}$, $M_{\mathrm{act}}^\star\gets M_{\mathrm{act}}$
    \Else
        \State \textbf{break}
    \EndIf
    \State $M_{\mathrm{act}}\gets M_{\mathrm{act}}+1$
\EndWhile
\State \Return $\boldsymbol{\alpha}^\star$, $M_{\mathrm{act}}^\star$
\end{algorithmic}
\end{algorithm}

\subsubsection{Nearest Chart-Point CSI Approximation (NN)}
This strategy uses CC embeddings to obtain approximate instantaneous CSI at the CPU and performs AAc by evaluating the EE objective in \eqref{eq:EE_problem} using the resulting channel approximation. In an offline calibration phase, each AP~$\ell$ collects local CSI realizations $\{\mathbf{h}^{(i)}_{\ell}\}$ from calibration users, computes the associated chart coordinates $\{\mathbf{z}^{(i)}_{\ell}\}$, and provides the resulting CSI-embedding pairs to the CPU as a lookup table. During online operation, given the embedding $\mathbf{z}_{k\ell}$ of user~$k$, the CPU retrieves the nearest calibration-user embedding following a CC-based nearest-neighbor CSI retrieval method similar to~\cite{10978279}, i.e.,
$i^\star_{k\ell}=\arg\min_i \|\mathbf{z}_{k\ell}-\mathbf{z}^{(i)}_\ell\|_2$,
and reuses the corresponding channel realization as
$\widehat{\mathbf{h}}_{k\ell}=\mathbf{h}^{(i^\star_{k\ell})}_{\ell}$.
Stacking $\{\widehat{\mathbf{h}}_{k\ell}\}$ over all users and APs yields the approximate global CSI matrix $\widehat{\mathbf{H}}$, and the antenna-activation vector is obtained by running Algorithm~\ref{alg:greedy_compact}. 
In this strategy, $\widehat{\mathbf{H}}$ is used both to compute the warm-start channel powers $u_{k,m}=|\widehat{h}_{k,m}|^2$ and to evaluate the objective $\mathrm{EE}(\boldsymbol{\alpha})$ during the greedy search.

\subsubsection{Decoder-Based CSI Reconstruction (RC)}
\label{subsubsec:decoder_recon}
Following the CC-based CSI compression and reconstruction framework in~\cite{abbass:hal-05393026}, a third strategy employs a neural encoder-decoder to reconstruct the instantaneous CSI at the CPU from low-dimensional CC embeddings, which is then used for AAc. 
For each served user~$k$, AP~$\ell$ compresses its locally estimated subband-averaged effective channel $\mathbf{h}_{k\ell}\in\mathbb{C}^{N}$ into an embedding $\mathbf{z}_{k\ell}\in\mathbb{R}^{d}$ and forwards it to the CPU. 
Using an AP-specific decoder $g_\ell(\cdot)$ trained offline as in~\cite{abbass:hal-05393026}, the CPU maps $\mathbf{z}_{k\ell}$ to a normalized channel direction $\widetilde{\mathbf{h}}_{k\ell}=g_\ell(\mathbf{z}_{k\ell})\in\mathbb{C}^{N}$ with $\|\widetilde{\mathbf{h}}_{k\ell}\|_2=1$. 
Since this normalized output preserves only the channel direction and discards the magnitude information, AP~$\ell$ additionally forwards the channel norm $\beta_{k\ell}=\|\mathbf{h}_{k\ell}\|_2$ as side information, allowing the CPU to restore the large-scale channel strength. In addition to this magnitude information, phase side information is required because CC encoding relies on phase-insensitive features, which may cause the reconstructed channels to exhibit AP-dependent phase ambiguities. Inspired by the phase-alignment mechanism in~\cite{abbass:hal-05393026}, we adopt a lightweight approximation tailored to the uniform cylindrical array with $N_{\mathrm{st}}$ vertical stages. Specifically, to provide a coarse phase reference for reconstruction, each AP computes a circular-mean phase $\phi_{k\ell,s}$ for each stage~$s$ from the locally estimated CSI and forwards these phase values as additional side information, i.e., $\{\phi_{k\ell,s}\}_{s=1}^{N_{\mathrm{st}}}$ along with $(\mathbf{z}_{k\ell},\beta_{k\ell})$. The CPU then reconstructs each antenna coefficient as
\(\widehat{h}_{k\ell,n}
=
\beta_{k\ell}\,\widetilde{h}_{k\ell,n}\,e^{j\phi_{k\ell,s(n)}}\).
Stacking the reconstructed vectors across all APs yields the global reconstructed CSI $\widehat{\mathbf{h}}_k\in\mathbb{C}^{M}$ for user~$k$. The corresponding user-wise vectors are then collected into $\widehat{\mathbf{H}}_{\mathrm{recon}}$, and the antenna-activation vector is obtained by running Algorithm~\ref{alg:greedy_compact}. In this strategy, $\widehat{\mathbf{H}}_{\mathrm{recon}}$ is used both to compute the warm-start channel powers $u_{k,m}=|\widehat{h}_{k,m}|^2$ and to evaluate the objective $\mathrm{EE}(\boldsymbol{\alpha})$ during the greedy search.

\subsection{Fronthaul Overhead and Computational Complexity}
Before the numerical results, we compare the proposed CC-based AAc strategies in terms of fronthaul signaling toward the CPU and online CSI/channel-power acquisition and EE-evaluation cost.
\subsubsection{Fronthaul Overhead Comparison}
Table~I reports the uplink fronthaul overhead per UE-AP link, measured as the number of transmitted real scalars, for AAc and uplink detection. Obtaining an AAc solution based on full instantaneous CSI at the CPU requires forwarding $2NS$ real scalars, whereas DD and NN require only the $d$-dimensional CC embedding, and RC requires $d+N_{\mathrm{st}}+1$ scalars due to the additional channel-norm and phase-alignment information, making DD and NN the most compact representations. For distributed detection, no additional CSI is exchanged beyond that required for AAc; hence, the total fronthaul coincides with the activation overhead, as reported in the ``AAc + Dist. Det.'' column. For centralized detection, AP~$\ell$ additionally forwards the CSI of its active antennas, adding $2N_{\ell}^{\mathrm{act}}S$ real scalars to the AAc signaling cost, where $N_{\ell}^{\mathrm{act}}$ is the number of active antennas at AP~$\ell$, as reported in the ``AAc + Cent. Det.'' column.
\begin{table}[t]
\centering
\textbf{TABLE I: Uplink fronthaul overhead per UE-AP link}\par\vspace{0.3em}
\label{tab:overhead_reform}
\scriptsize
\setlength{\tabcolsep}{3.5pt}
\renewcommand{\arraystretch}{1.05}
\begin{tabular}{lccc}
\toprule
\textbf{Method} 
& \textbf{AAc Overhead} 
& \textbf{Total Overhead} 
& \textbf{Total Overhead} \\
& 
& \textbf{(AAc + Dist. Det.)} 
& \textbf{(AAc + Cent. Det.)} \\
\midrule
Full CSI 
& $2NS$ 
& $2NS$ 
& $2NS$ \\
\midrule
DD 
& $d$ 
& $d$ 
& $d+2N_{\ell}^{\mathrm{act}}S$ \\
NN 
& $d$ 
& $d$ 
& $d+2N_{\ell}^{\mathrm{act}}S$ \\
RC 
& $d+N_{\mathrm{st}}+1$ 
& $d+N_{\mathrm{st}}+1$ 
& $d+N_{\mathrm{st}}+1+2N_{\ell}^{\mathrm{act}}S$ \\
\bottomrule
\end{tabular}
\end{table}

\subsubsection{Computational Complexity Analysis}
We compare the AAc strategies in terms of CSI/channel-power acquisition and objective-evaluation cost, as summarized in Table~II.
DD and NN rely on a CPU-stored lookup table and perform nearest-neighbor search over $N_{\mathrm{c}}$ calibration embeddings per AP, yielding an online acquisition complexity of $\mathcal{O}(LK N_{\mathrm{c}} d)$; this step predicts per-antenna channel-power vectors for DD and retrieves approximate CSI for NN.
RC avoids lookup-table search by reconstructing CSI with a trained decoder, with online cost $\mathcal{O}\!\left(LK(FT+T^{2}+TN)\right)$, where $F$ and $T$ are decoder parameters denoting the number of random Fourier features and the hidden-layer width, respectively.
The DD strategy evaluates a CC-based proxy EE with complexity $\mathcal{O}(K M_{\mathrm{act}})$ per objective evaluation, while NN and RC rely on the exact MMSE-based EE, whose dominant cost is the MMSE covariance-matrix inversion, scaling as $\mathcal{O}(K M_{\mathrm{act}}^{3})$ per objective evaluation.
Overall, DD has the lowest total online computational complexity among the three strategies, owing to its embedding-domain operation and low-complexity proxy-EE evaluation.
\begin{table}[t]
\centering
\textbf{TABLE II: Complexity summary of the proposed AAc strategies}\par\vspace{0.3em}
\label{tab:complexity_summary}
\scriptsize
\setlength{\tabcolsep}{3.5pt}
\renewcommand{\arraystretch}{1.1}
\begin{tabular}{lcc}
\toprule
\textbf{Method} 
& \textbf{CSI/Channel-Power Acquisition} 
& \textbf{Objective Evaluation} \\
\midrule
DD 
& $\mathcal{O}(LK N_{\mathrm{c}} d)$ 
& $\mathcal{O}(K M_{\mathrm{act}})$ \\

NN 
& $\mathcal{O}(LK N_{\mathrm{c}} d)$ 
& $\mathcal{O}(K M_{\mathrm{act}}^{3})$ \\

RC 
& $\mathcal{O}\!\left(LK(FT+T^{2}+TN)\right)$ 
& $\mathcal{O}(K M_{\mathrm{act}}^{3})$ \\
\bottomrule
\end{tabular}
\end{table}

\section{Performance Evaluation}
This section evaluates the DD, NN, and RC strategies. These methods are compared with ground-truth (GT) benchmarks, corresponding to solutions obtained under the availability of full instantaneous CSI at the CPU. Each curve is labeled following the convention \texttt{strategy\,-\,combiner\,-\,algorithm}, where the strategy is one of the proposed schemes (DD, NN, RC) or the GT benchmark, the combiner is centralized (C\_MMSE) or distributed (D\_MMSE) MMSE detection, and the algorithm indicates how the activation vector is obtained: greedy (G), simulated annealing (SA), or all-active (AA). The greedy algorithm (G) refers to Algorithm~\ref{alg:greedy_compact}; when paired with the GT strategy, it operates on full CSI at the CPU. Simulated annealing (SA), implemented following~\cite{delahaye2019sa}, uses a search procedure different from the proposed greedy algorithm and directly evaluates the exact EE objective. The all-active case (AA) keeps all antennas active, applying no activation-search algorithm. The remainder of this section presents the simulation setup and evaluates the proposed strategies and benchmarks in terms of EE and sum spectral efficiency (SE), concluding with a summary of the resulting performance trends.

\textbf{Simulation Setup:}
The proposed framework is evaluated using realistic datasets generated with \textit{Sionna}~\cite{sionna} for the \emph{Étoile} urban scenario in Paris, with the main parameters summarized in Table III.
For DD and NN, $N_{\mathrm{c}}=8000$ calibration realizations are used for CC construction and $2000$ realizations for testing, while RC uses $N_{\mathrm{c}}=3500$, $4500$, and $2000$ realizations for CC calibration, autoencoder training, and testing, respectively.

\begin{table}[t]
\centering
\textbf{TABLE III: Simulation Parameters}\par\vspace{0.3em}
\label{tab:sim_setup}
\scriptsize
\setlength{\tabcolsep}{2.0pt}
\renewcommand{\arraystretch}{1.08}
\begin{tabular}{l c l c}
\toprule
\textbf{Param.} & \textbf{Value} & \textbf{Param.} & \textbf{Value} \\
\midrule
$(f_c,B_{\mathrm{sub}})$ [MHz] & $(3800,\,0.36)$ &
$(S,\tau_p)$ & $(12,\,15)$ \\
$(L,N,K,N_{\mathrm{st}})$ & $(10,\,48,\,30,\,4)$ &
$d$ & $4$ \\
$(p_k,\sigma^2)$ [dBm] & $(10,\,-93)$ &
$\nu_k$ & $0.4$ \\
$(P_{\mathrm{fix}},P_{\mathrm{tc}})$ [W] & $(0.825,\,0.2)$ &
$P_{\mathrm{ft}}$ [W/bps] & $0.25{\times}10^{-9}$ \\
$F$ & $200$ &
$T$ & $128$ \\
\bottomrule
\end{tabular}
\end{table}

\begin{figure}[t]
\centering
\includegraphics[width=0.95\columnwidth]{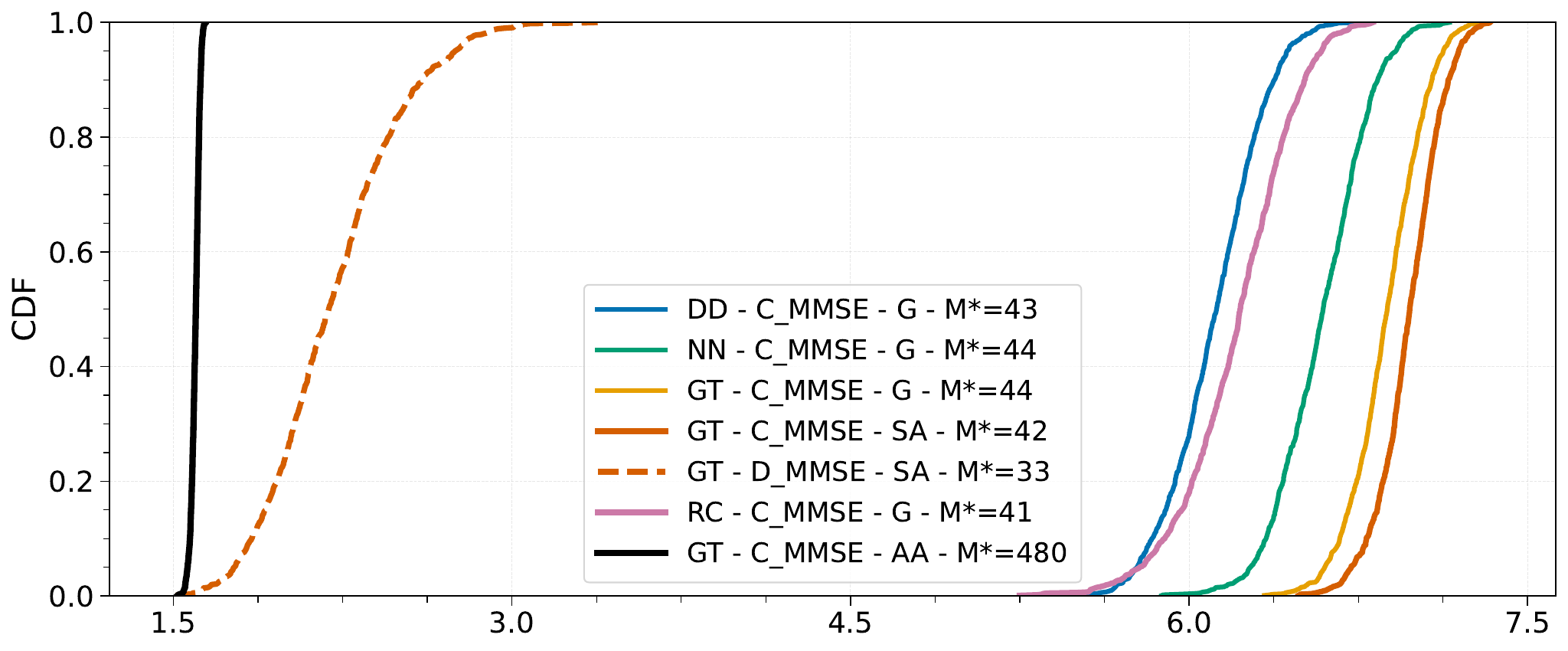}

\includegraphics[width=0.95\columnwidth]{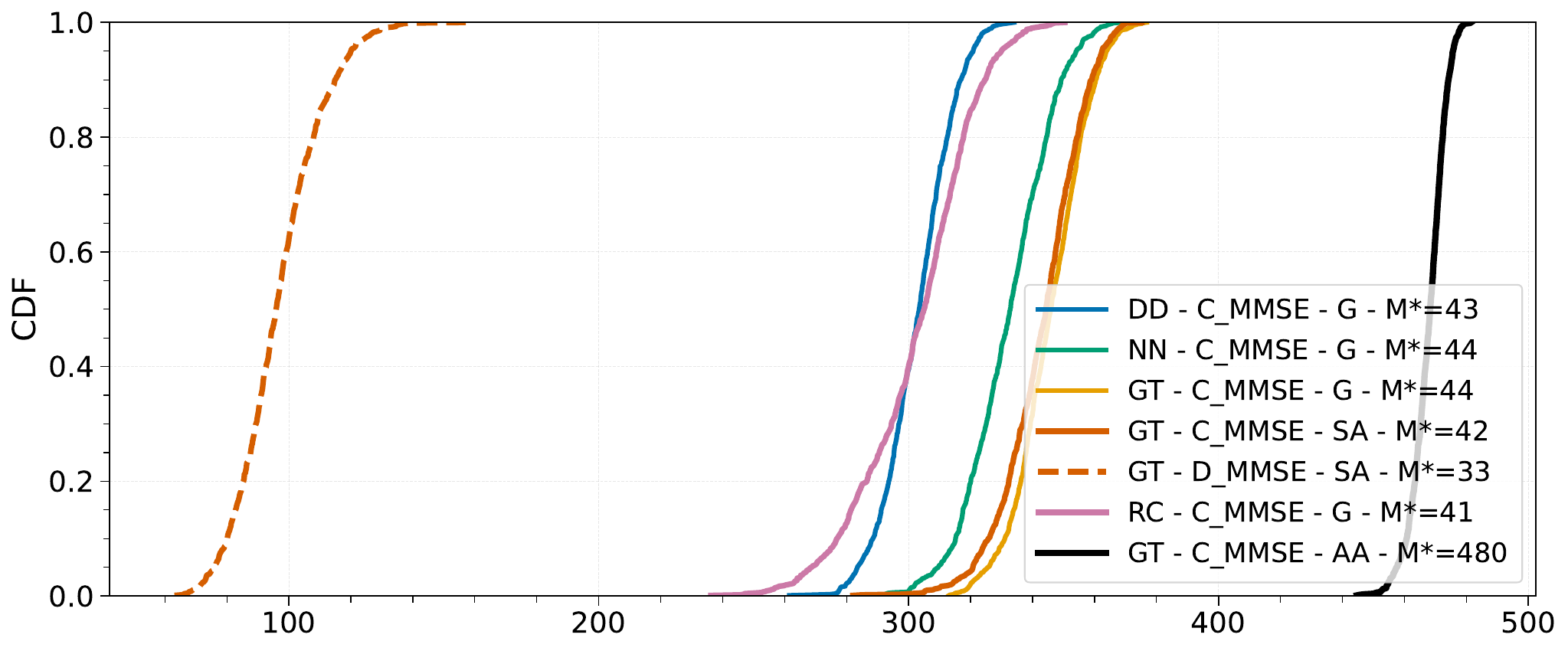}

\caption{CDFs of (top) EE [Mbit/Joule] and (bottom) Sum spectral efficiency [bit/s/Hz] for the considered AAc strategies under MMSE reception.}
\label{fig:ee_se_cdf}
\end{figure}

\textbf{Performance comparisons:}
Fig.~\ref{fig:ee_se_cdf} shows the empirical CDFs of EE and sum SE over $1000$ Monte-Carlo realizations. The all-active baseline GT-C\_MMSE-AA achieves the highest SE but the lowest EE, due to excessive circuit and processing power consumption, resulting in a left-shifted EE curve with limited variability. In contrast, the evaluated AAc schemes substantially improve EE, shifting the EE CDFs to the right while incurring only moderate SE losses. The proposed CC-driven greedy methods, DD-C\_MMSE-G, NN-C\_MMSE-G, and RC-C\_MMSE-G, achieve average EE gains of approximately $280$-$315\%$ over GT-C\_MMSE-AA while activating only $40$-$45$ antennas out of $M=480$. Among them, NN-C\_MMSE-G is closest to the full-CSI-based GT benchmarks, whereas DD-C\_MMSE-G and RC-C\_MMSE-G are slightly degraded by proxy-based optimization and decoder reconstruction loss, respectively. GT-C\_MMSE-SA achieves the highest observed EE, while GT-C\_MMSE-G yields a nearly identical CDF at significantly lower complexity. In our implementation, SA performs $N_{\mathrm{SA}}=1000$ candidate-swap trials per tested antenna cardinality, each requiring one exact EE evaluation with complexity $\mathcal{O}(K M_{\mathrm{act}}^{3})$; further implementation details are omitted due to space limitations, and a general description of SA can be found in~\cite{delahaye2019sa}. By contrast, GT-D\_MMSE-SA shows a pronounced EE and SE degradation compared with centralized detection, confirming that centralized detection is important for high performance, while the proposed CC-driven strategies aim to make it scalable by reducing CSI signaling overhead.

\textbf{Trade-off summary:}
Fig.~\ref{fig:tradeoff_ranking} summarizes the performance-overhead-complexity trade-offs among the proposed AAc strategies. NN achieves the best EE performance and the strongest performance-overhead compromise, making it the preferred strategy under the considered setup. DD provides the lowest signaling and computational cost at the price of lower EE accuracy, whereas RC improves over DD in terms of EE but requires additional side information and higher decoder-based reconstruction complexity.

\begin{figure}[t]
\centering
\begin{tikzpicture}[font=\scriptsize, x=0.72cm, y=0.45cm]

\def\xlab{-0.2}
\def\xone{2.0}
\def\xtwo{3.1}
\def\xthree{4.2}
\def\xfour{5.3}
\def\xfive{6.4}

\node[anchor=east] at (\xlab,3) {\textbf{EE performance}};
\node[anchor=east] at (\xlab,2) {\textbf{Signaling efficiency}};
\node[anchor=east] at (\xlab,1) {\textbf{Computational efficiency}};

\node at (\xone,3) {\textbf{DD}};
\node at (\xtwo,3) {$<$};
\node at (\xthree,3) {\textbf{RC}};
\node at (\xfour,3) {$<$};
\node at (\xfive,3) {\textbf{NN}};

\node at (\xone,2) {\textbf{RC}};
\node at (\xtwo,2) {$<$};
\node at (\xthree,2) {\textbf{DD}};
\node at (\xfour,2) {$=$};
\node at (\xfive,2) {\textbf{NN}};

\node at (\xone,1) {\textbf{RC}};
\node at (\xtwo,1) {$<$};
\node at (\xthree,1) {\textbf{NN}};
\node at (\xfour,1) {$<$};
\node at (\xfive,1) {\textbf{DD}};

\draw[->, thick] (1.6,0.35) -- (6.8,0.35);
\node[anchor=north] at (4.2,0.25) {better};

\end{tikzpicture}
\vspace{-1em}
\caption{Qualitative EE-signaling-complexity trade-off among the proposed CC-driven AAc strategies.}
\label{fig:tradeoff_ranking}
\end{figure}

\section{Conclusion}
This work demonstrated that channel charting enables scalable centralized antenna activation in CF-mMIMO systems by replacing full instantaneous CSI exchange with compact CC embeddings. Three CC-based strategies--DD, NN, and RC--were proposed to exploit these embeddings for network-wide EE optimization under different levels of CSI approximation, revealing clear trade-offs between performance, signaling, and computational complexity. The results, obtained using realistic antenna radiation patterns in a ray-tracing environment, show that the proposed strategies achieve competitive EE performance with respect to the considered full-CSI benchmarks while requiring substantially lower signaling. These findings confirm that CC-based representations can successfully support centralized energy-efficient antenna activation, with the preferred strategy depending on whether the main priority is signaling overhead, online complexity, or EE performance.

\textbf{Acknowledgment: }This work was supported by the ANR under the France~2030 program, grant ``NF-PERSEUS: ANR-22-PEFT-0004''.
\bibliographystyle{IEEEtran}
\bibliography{References}
\end{document}